\documentstyle[aps,graphicx,amsmath,amssymb,bm]{revtex}
\begin{document}
\newcommand{\be}{\begin{equation}}
\newcommand{\ee}{\end{equation}}
\newcommand{\p}{\partial}
\newcommand{\hp}{\widehat{\p}}
\newcommand{\ov}{\overline}
\newcommand{\da}{^{\dagger}}
\newcommand{\w}{\wedge}
\newcommand{\st}{\stackrel}
\newcommand{\mb}{\mbox}
\newcommand{\mx}{\mbox}
\newcommand{\mt}{\mathtt}
\newcommand{\dt}{\mathtt{d}}
\newcommand{\al}{\alpha}
\newcommand{\bb}{\beta}
\newcommand{\ga}{\gamma}
\newcommand{\te}{\theta}
\newcommand{\Te}{\Theta}
\newcommand{\de}{\delta}
\newcommand{\et}{\tilde{e}}
\newcommand{\ze}{\xi}
\newcommand{\s}{\sigma}
\newcommand{\e}{\epsilon}
\newcommand{\om}{\omega}
\newcommand{\Om}{\Omega}
\newcommand{\la}{\lambda}
\newcommand{\La}{\Lambda}
\newcommand{\n}{\nabla}
\newcommand{\hn}{\widehat{\nabla}}
\newcommand{\hph}{\widehat{\phi}}
\newcommand{\ah}{\widehat{a}}
\newcommand{\bh}{\widehat{b}}
\newcommand{\ch}{\widehat{c}}
\newcommand{\ddh}{\widehat{d}}
\newcommand{\eh}{\widehat{e}}
\newcommand{\gh}{\widehat{g}}
\newcommand{\ph}{\widehat{p}}
\newcommand{\qh}{\widehat{q}}
\newcommand{\mh}{\widehat{m}}
\newcommand{\nh}{\widehat{n}}
\newcommand{\Dh}{\widehat{D}}
\newcommand{\stu}{\st{\sqcup}}
\newcommand{\au}{\stu{a}}
\newcommand{\bu}{\stu{b}}
\newcommand{\cu}{\stu{c}}
\newcommand{\du}{\stu{d}}
\newcommand{\eu}{\stu{e}}
\newcommand{\mmu}{\stu{m}}
\newcommand{\nnu}{\stu{n}}
\newcommand{\pu}{\stu{p}}
\newcommand{\Du}{\stu{D}}
\newcommand{\sto}{\st{\odot}}
\newcommand{\as}{\st{\odot}{a}}
\newcommand{\bs}{\st{\odot}{b}}
\newcommand{\cs}{\st{\odot}{c}}
\newcommand{\ds}{\st{\odot}{d}}
\newcommand{\es}{\st{\odot}{e}}
\newcommand{\gs}{\st{\odot}{g}}
\newcommand{\ms}{\st{\odot}{m}}
\newcommand{\ns}{\st{\odot}{n}}
\newcommand{\ps}{\st{\odot}{p}}
\newcommand{\Ds}{\st{\odot}{D}}
\newcommand{\sts}{\st{\spadesuit}}
\newcommand{\sth}{\st{\heartsuit}}
\newcommand{\stp}{\st{\perp}}
\newcommand{\std}{\st{\diamondsuit}}
\newcommand{\ad}{\st{\spadesuit}{a}}
\newcommand{\bd}{\st{\spadesuit}{b}}
\newcommand{\cd}{\st{\spadesuit}{c}}
\newcommand{\gd}{\st{\spadesuit}{g}}
\newcommand{\dd}{\st{\spadesuit}{d}}
\newcommand{\Dd}{\st{\spadesuit}{D}}
\newcommand{\ed}{\st{\spadesuit}{e}}
\newcommand{\fd}{\st{\spadesuit}{f}}
\newcommand{\zd}{\st{\spadesuit}{\xi}}
\newcommand{\md}{\st{\spadesuit}{m}}
\newcommand{\nd}{\st{\spadesuit}{n}}
\newcommand{\stc}{\st{\clubsuit}}
\newcommand{\az}{\st{\clubsuit}{a}}
\newcommand{\bz}{\st{\clubsuit}{b}}
\newcommand{\cz}{\st{\clubsuit}{c}}
\newcommand{\dz}{\st{\clubsuit}{d}}
\newcommand{\Dz}{\st{\clubsuit}{D}}
\newcommand{\ez}{\st{\clubsuit}{e}}
\newcommand{\fz}{\st{\clubsuit}{f}}
\newcommand{\nz}{\st{\clubsuit}{n}}
\newcommand{\mz}{\st{\clubsuit}{m}}
\newcommand{\tb}{\overline{\theta}}
\newcommand{\ti}{\widetilde}

\newcommand{\2}{\frac{1}{2}}
\newcommand{\3}{\frac{1}{3}}
\newcommand{\4}{\frac{1}{4}}
\newcommand{\8}{\frac{1}{8}}
\newcommand{\6}{\frac{1}{16}}

\newcommand{\ra}{\rightarrow}
\newcommand{\Ra}{\Rightarrow}
\newcommand{\im}{\Longleftrightarrow}
\newcommand{\vs}{\vspace{5mm}}
\newcommand{\x}{\star}
\newcommand{\Delt}{\p^{\star}}

\thispagestyle{empty}
{\bf 9th November, 2003} \hspace{\fill}

\vspace{1cm}
\begin{center}{\Large{\bf COSMOLOGY WITH BRANES WRAPPING\\

\vspace{3mm}
CURVED INTERNAL MANIFOLDS}}\\
\vspace{1cm}
{\large{\bf T. Biswas\footnote{tirtho@hep.physics.mcgill.ca} }}\\
\vspace{5mm}
{\small Center for High Energy Physics\\
Department of Physics\\
McGill University\\
Montreal, Quebec H3A 2T8}
\end{center}

\begin{abstract}
In this paper we first derive solutions which can be interpreted as branes
wrapping nontrivial curved manifolds, and then study
their cosmological implications. We find that at early times the branes tend
to shrink the internal manifold, while allowing the ``unwrapped'' dimensions
to expand in congruence with what has already been observed in the case when
the internal manifold is flat (tori). However, at late times the internal
curvature terms become important leading to potentially interesting
differences.
\end {abstract}

\newpage
\setcounter{page}{1}
\begin{center}  {\large  {\bf INTRODUCTION} }
\end{center}
Inspired by String/M-Theory brane-physics have blossomed in recent years. 
Branes are $p$ dimensional extended objects embedded in a higher $\Dh \geq 
p+1$ dimensional universe which arise naturally in String theory  as 
hyperplanes where open strings can end, and in Supergravity/M-Threory as 
solitons \cite{argurio}. Mostly, branes have found applications in two 
virtually opposite set ups: In the ``brane-world'' scenario \cite{randall} 
the brane dimensions coincide with the three observed dimensions of our 
universe, while the spatial dimensions perpendicular to the brane, which can 
be both compact or non-compact correspond to the ``unseen internal 
dimensions''. Contrastingly in the ``brane-gas'' scenario \cite{bgas,wafa} 
the branes wrap around compact internal dimensions, while the directions  
perpendicular to the branes become the observed dimensions of our universe. 
While the brane-world picture have several virtues including being able to 
address the hierarchy  and the cosmological constant problem, the brane gas 
model  can explain why the internal manifold remained small 
as compared to the observed dimensions, and a simple counting argument also 
yields the dimensionality of the observed universe to be three! Invoking 
T-duality brane gas cosmology (BGC) also seems to be able to avoid the big 
bang singularity. Thus, it may be an interesting venture to combine the two 
scenarios by considering branes  some (three) of whose dimensions are   
noncompact (which becomes our observed universe) while the others  wrap 
compact internal manifolds,  which potentially can be non-trivially curved - 
the isometries of this curved internal manifold can then be associated with 
gauge fields living on the brane. In this paper however,  we only restrict 
ourselves to the usual set up of BGC\footnote{In the strict sense our study  is slightly different from BGC as the dilaton is fixed to be a constant in our case as we will soon find out. However, the analysis is related and the value of the dilaton is supposed to get stabilized at later times any way.} but we generalize the internal manifold from being a flat tori to a curved space.

In \cite{wafa} Brandenberger and Vafa observed that string winding states 
would tend to prevent expansion since the energy of the states increases if 
the circumference increases. However, strings with opposite winding numbers 
can annihilate if their world volume intersects and this happens most 
efficiently in three dimensions. Thus in a three dimensional subspace the 
string winding states can annihilate letting these dimensions grow, while  
strings winding other directions will not be able to annihilate each other 
efficiently and ultimately fall out of thermal equilibrium thereby stopping 
the expansion. This  stabilization mechanism of the internal manifold by employing `string gas' was further quantified in \cite{watson} and has also been generalized to `brane gas' 
scenarios \cite{brandenberger}  (for more details also see \cite{kabat}) with essentially  similar results, except 
that now we have a hierarchy in the sizes of the extra dimensions, coming 
from the contributions of the p-branes, with different $p$, as in general 
$p$-branes annihilate most effectively in $2p+1$ dimensions. Although BGC 
has been successful in solving several problems associated with standard big 
bang cosmology, there are some unresolved issues as pointed out  in 
\cite{kaya}. For example, it is not clear exactly how the three dimensions 
that we observe today are chosen among all the other dimensions, perhaps 
through random quantum and thermal fluctuations. In that case it is possible 
that our universe may contain different patches where different directions 
have become large! Also, it is known that branes with different values of 
$p$ can interact with each other, changing their winding numbers, thereby 
allowing thermal equilibrium to be maintained among the brane winding 
states. A simple way to eliminate these issues is to assume the universe to 
be a direct product of three non-compact dimensions and a compact internal 
manifold. The BGC in such a set up has been extensively studied in 
\cite{kaya} when the internal manifold is a tori. Here, we adopt the same 
approach, but generalize the program to nontrivial curved internal manifolds\footnote{It should be mentioned that BGC with branes wrapping nontrivial
cycles (in Calabi-Yau  and K3 spaces)  was first studied in \cite{curved}.}  
  to study how, if at all, the internal curvature may change the 
cosmological dynamics. Of course, in this set up the question of why the 
universe is a product space with specifically three non-compact dimensions 
is an open issue, and the nice  BGC argument   concerning the dimensionality 
of our observed universe becomes redundant. We will come back to this at the 
end.

For a flat $p$-brane the $p$ longitudinal (along the brane) directions are 
flat, or in other words a flat $p$-brane solution preserves the isometries 
of the $p$ dimensional flat space. These branes can then wrap around flat 
$p$-dimensional tori. If one now considers branes wrapping a curved internal 
manifold, say $M_p$ then the brane solution should accordingly, preserve the 
isometries of $M_p$. In section I, we derive such $p$-brane solutions\footnote{Previously branes wrapping curved dimensions have  been 
studied using boundary state conformal field theory \cite{schomerus}, but 
here we take a supergravity approach.} which 
look like de Sitter black holes from the four dimensional point of 
view. This is to be expected as after 
reduction ordinary $p$-branes  resemble black hole solutions, and since our 
internal background manifold is curved, we are living in an asymptotically 
de Sitter space-time. We here only obtain the uncharged solution but it may 
be interesting to study other charged solutions which will  have underlying 
connections to non-abelian black holes (see \cite{non-abelian} for a 
review).  In section II, we obtain  the back reaction  of these branes on 
the space-time geometry by analogy with the flat brane solution \cite{kaya} 
and check that the stress energy tensor is consistent.  As is usual the 
branes represent delta function sources for the Einstein's equations which 
can be smoothed out by considering a uniform density of brane gas. In  
section III, we study BGC both without and with matter/radiation. We find 
that the early evolution resembles the picture with flat branes 
\cite{wafa,brandenberger,kaya}: while the branes wrapping the internal 
manifold tends to contract them, the internal curvature terms only 
reinforcing this effect, the noncompact dimensions on the other hand are 
free to expand as the branes act as pressureless dust. At late times however 
the picture changes as the curvature term starts dominating over the brane 
effects. The stabilization of the internal manifold that was achieved in the 
presence of ordinary matter and branes (the expansion of ordinary matter 
being balanced by the contraction of the branes) is lost. Preliminary  
analysis indicates that the internal manifold may also be  growing along 
with the ordinary dimensions. This suggests that we may have to invoke other 
mechanisms to stabilize the internal manifold at late times, like turning on 
the fluxes \cite{flux}. We conclude by summarizing and commenting on 
possible future research.

\begin{center} {\bf {\large I. BRANES WRAPPING CURVED MANIFOLDS }} 
\end{center}
{\bf Supergravity and Brane Ansatz:} We start with a generic bosonic 
sector\footnote{In brane solutions all fermionic fields are typically put to 
zero.} of a supergravity (SUGRA) action:
\begin{equation}
\widehat{S}=\frac{1}{16\pi G_{\Dh}}\int 
dx^{\Dh}\sqrt{-g}\{\widehat{R}-\2\p_{\mh}\phi\p^{\mh}\phi-\2\sum_I 
\frac{1}{n_I!}e^{a_I\phi}F^2_{n_I}-2\widehat{\La}\}
\label{eq:sugra}
\end{equation}
where $\phi$ is the dilaton field and field strengths $F_{n_I}$'s are  $n_I$ 
  forms, $I=1\dots M$ and we have also included a cosmological constant 
term. Although both in the action and in the field strengths one can have 
Chern-Simons-like terms it is known that they can be ignored while 
considering brane like solutions. The field equations that follow from 
(\ref{eq:sugra}) are
\be
G_{\mh\nh}-\2(\p_{\mh}\phi\p_{\nh}\phi-\2 g_{\mh\nh}(\p\phi)^2)-\2\sum_I 
\frac{1}{n_I!}e^{a_I\phi}(n_IF_{\mh\ph_2\dots\ph_I}F_{\nh}{}^{\ph_2\dots\ph_I}-\2 
g_{\mh\nh}F^2_{n_I})+\widehat{\La}g_{\mh\nh}=0
\label{eq:einstein}
\ee
\be
\frac{1}{\sqrt{-g}}\p_{\mh}(\sqrt{-g}g^{\mh\nh}\p_{\nh}\phi)-\2\sum_I 
\frac{1}{n_I!}e^{a_I\phi}F^2_{n_I}=0
\label{eq:gauge}
\ee
\begin{equation}
\frac{1}{(n-1)!}\frac{1}{\sqrt{-g}}\p_{\mh}(\sqrt{-g}e^{a_I\phi}F^{\mh\ph_2\dots\ph_I})=0
\label{eq:dilaton}
\end{equation}
We now specialize to the uncharged solution, for which we can consistently 
put the field-strength and the dilaton to zero. We are thus left with only 
(\ref{eq:einstein})\footnote{We are essentially in Einstein's  gravity with 
a cosmological constant.}; (\ref{eq:gauge}) and (\ref{eq:dilaton}) being 
trivially satisfied.

Now let us look at the ansatz for our $p$-brane metric. For simplicity we 
assume the internal manifold to be a group manifold $G_{\Ds}$, but one 
should be able to generalize the results to other homogeneous spaces. Since 
in our picture the $p$-brane is wrapping the internal manifold, $\Ds=p$. 
Moreover, the $p$-brane metric have to possess the isometries of the 
internal group manifold. As in the flat $p$-brane case, this fixes the $\Ds$ 
directional components of the metric:
\begin{equation}
g_{\mh\nh}(x,y)=\left( \begin{array}{cc}
g_{mn}(x) & 0\\
0 & \Psi(x)g^K_{\ms\ns}(y)
\end{array} \right)
\end{equation}
We employ symbols `$\odot$' and `no symbol' to denote quantities 
corresponding to the internal and the external manifold respectively, 
wherever necessary. Also, $x$ and $y$ denote coordinates charting the 
observational universe and the internal manifold respectively. Here $g^K$ is the 
Killing metric\footnote{One can also employ more general ``squashed metric'' 
but we will not consider them in this manuscript.} possessing the isometries 
of the group manifold and $\Psi(x)$ is the radion.  If we now further impose 
invariance under time translations and $SO(D)$ rotations among the 
transverse $x$ directions then we obtain
\be
\widehat{d}s^2=-\widehat{\beta}^2dt^2+\widehat{\Phi}^2dr^2+\widehat{\Gamma}^2r^2d\Om^2_{D-1}+\Psi^2\ds 
s^2
\label{eq:metric}
\ee
where all the functions only depend on the radial coordinate $r$, the brane 
being located at $r=0$. Note that the only way this metric differs from the 
usual flat brane ansatz is that the metric along the brane directions is not 
flat but corresponds to the Killing metric of the group manifold. At this 
point one could use (\ref{eq:metric}) to solve the field equations 
(\ref{eq:einstein}) but it is easier to perform a dimensional reduction 
which then maps the $p$-branes to black holes. Since the internal manifold 
is curved we, in fact, expect the $p$-branes to look like de Sitter black 
holes \cite{gibbons}. \vs\\
{\bf Reduction and Brane Solution:} To perform the reduction it is 
convenient to work with the vielbein:
\begin{equation}
\eh_{\mh}{}^{\ah}=\left( \begin{array}{cc}
e_m{}^{a}(x) & 0\\
0 &\Psi(x)\es_{\ms}{}^{\as}(y)
\end{array} \right);
\hspace{5mm}
\eh_{\ah}{}^{\mh}=\left( \begin{array}{cc}
e_{a}{}^{m}(x) & 0\\
0 &\Psi^{-1}(x)\es_{\as}{}^{\ms}(y)
\end{array} \right)
\end{equation}
and the ``flat-metric''
\begin{equation}
g_{\ah\bh}=\left( \begin{array}{cc}
\eta_{ab} & 0\\
0 & g^K_{\as\bs}
\end{array} \right)
\label{eq:flatmetric}
\end{equation}
Such a dimensional reduction is known to be consistent and yields
$$\widehat{S}=\frac{1}{16\pi G_{\Dh}}\int 
dx^{\Dh}\sqrt{-\gh}(\widehat{R}-2\widehat{\Lambda})=\frac{1}{16\pi G_{\Dh}}\int 
dy^{\Ds}\sqrt{-\gs}\int dx^{D+1}\sqrt{-g}\Psi^{\Ds}[R+K-V]$$
$$=\frac{1}{16\pi G_{D+1}}\int dx^{D+1}\sqrt{-g}\Psi^{\Ds}[R+K-V]$$
where
$$K=-(2\Ds\frac{\Box \Psi}{\Psi}+\Ds(\Ds-1)\frac{(\p\Psi)^2}{\Psi^2})\ ;\ 
V=-\frac{\Ds}{4}\Psi^{-2}+2\widehat{\Lambda}$$
and
$$\frac{1}{16\pi G_{\Dh}}\int dy^{\Ds}\sqrt{-\gs}=\frac{1}{16\pi 
G_{\Dh}}V_G=\frac{1}{16\pi G_{D+1}}$$

To obtain the action in the Einstein frame, we perform the well known 
conformal re-scalings
\be
\eh_{\ah}{}^{\mh}=\Delta\eh'_{\ah}{}^{\mh}\ ;\ 
\Delta=\Psi'^{\frac{\Ds}{\Dh-2}}
\ee
Then we have
$$ S=\frac{1}{16\pi G_{D+1}}\int dx^{D+1}e'^{-1}\left[R'+K'-V'\right]$$
with
\be
K'=-\frac{\Ds(D-1)}{\Dh-2}(\p' \psi)^2\ ;\ 
V'=2\widehat{\Lambda}e^{-2\frac{\Ds}{\Dh-2}\psi}-\frac{\Ds}{4}e^{-2\psi}
\label{eq:4daction}
\ee
where we have defined
$$\Psi'=e^{\psi}$$
In future we will drop the `primes'.

The action (\ref{eq:4daction}) yields field equations for the four 
dimensional metric and the radion. However, we know that for uncharged 
$p$-branes the radion is a constant  satisfying
$$\frac{\p V(\psi)}{\p \psi}=0$$
\be
\Ra e^{2\psi}=\left(\frac{\Dh-2}{8\widehat{\Lambda}}\right)^{\frac{\Dh-2}{D-1}}
\ee
The potential $V$ then acts as an effective $D+1$ dimensional cosmological 
constant $\Lambda$ given by
\be
\La\equiv\frac{D-1}{8}e^{-2\psi}=\frac{D-1}{8}\left(\frac{8\widehat{\Lambda}}{\Dh-2}\right)^{\frac{\Dh-2}{D-1}}
\ee

The effective action now reads
\be
S=\frac{1}{16\pi G_{D+1}}\int dx^{D+1}e^{-1}\left[R-2\La\right]
\label{eq:dsaction}
\ee
De Sitter black hole solutions for (\ref{eq:dsaction}) are known \cite{gibbons}:
\be
ds^2=-\beta^2dt^2+\Phi^2dr^2+\Gamma^2r^2d\Om^2_{D-1}
\ee
with
\be
\Gamma=1\ ;\ \beta=\Phi^{-1}\ ;\ 
\Phi^2=\frac{1}{1-\frac{T_p}{r^{D-2}}-\frac{2\La}{D(D-1)}r^2}\equiv 
\frac{1}{f(r)}
\ee
The ADM mass of the black hole $T_p$, arises  as an integration constant and 
is equivalent to the ADM mass or tension  of the $p$-brane \cite{mass}. For 
completeness sake, let us write down the full brane solution
\be
\widehat{d}s^2=\left(\frac{8\widehat{\Lambda}}{\Dh-2}\right)^{\frac{\Ds}{D-1}}\left(-f(r)dt^2+\frac{1}{f(r)}dr^2+r^2d\Om^2_{D-1}+\left(\frac{\Dh-2}{8\widehat{\Lambda}}\right)^{\frac{\Dh-2}{D-1}}\ds 
s^2\right)
\ee
\pagebreak
\begin{center} {\bf {\large II. BRANE BACK REACTION}}
\end{center}
Having obtained the brane solution wrapping the extra dimensional manifold, 
the next task would be to compute its back reaction on the background 
geometry. The cosmological ansatz for the background metric looks like
\be
ds^2=-\widehat{\beta}^2dt^2+\widehat{\alpha}^2dx^2+\Psi^2\ds y^2
\ee
where we take the external three dimensional spatial geometry to   be flat.  
Again it is useful to introduce the vielbeins
\begin{equation}
\eh_{\ah}{}^{\mh}=\left( \begin{array}{ccc}
e^{-W} & 0 & 0\\
0 &e^{-A}\de_{\au}{}^{\mmu}& 0\\
0 & 0 & e^{-S}\es_{\as}{}^{\ms}
\end{array} \right)
\label{eq:vielbein}
\end{equation}
along with the flat metric (\ref{eq:flatmetric}). We use the symbol `$\sqcup$' for quantities and indices corresponding to the external spatial dimensions. Here we have also 
redefined  $\beta,\alpha$ and $\Psi$ as exponentials of $W$, $A$ and $S$ 
respectively, which are now to be treated as collective coordinates 
characterizing the geometry of both the internal and the external manifold, 
depending  only on time. In \cite{kaya} the stress energy tensor for a flat 
brane located at $x_0$ in such a background was already obtained which lends 
a natural generalization to curved branes\footnote{Although the result 
physically makes sense because we expect the branes when wrapped around the 
internal manifold (curved or flat) to behave as delta function sources in 
the transverse (observable) directions, it would be nice to derive it ab 
initio starting from, for example, the Polyakov brane action. This however 
would require a better understanding of the brane dynamics which may also 
lead to some corrections to (\ref{eq:stress}).}:
$$ T_{00}=T_pe^{-DA}\de(x-x_0)$$
$$T_{\au\bu}=0$$
\be
T_{\as\bs}=-T_pe^{-DA}\de(x-x_0)g_{\as\bs}
\label{eq:stress}
\ee
Consider now the brane gas scenario with say $n_l$ branes located at $x_l$ 
which implies $\de(x-x_0)\ra \sum_{x_l} n_l\de(x-x_l)$. As is standard 
practice we now pass on from a discrete to  a continuous distribution of 
brane gas
$$\sum_{x_l} n_l\de(x-x_l)\ra \int dx' n(x')\de(x-x')$$
where $n(x)$ is the density of brane gas. Assuming a uniform density then 
gives us
$$ T_{00}=nT_pe^{-DA}$$
$$T_{\au\bu}=0$$
\be
T_{\as\bs}=-nT_pe^{-DA}g_{\as\bs}
\label{eq:branestress}
\ee
This stress-energy tensor differs from the usual brane gas stress-energy tensor 
(see for example \cite{brandenberger}) which has an equation of state 
parameter $-p/(\Dh-1)$ along all dimensions. As we will see later, 
(\ref{eq:branestress}) corresponds to  equation of state parameters -1 along 
the wrapped compact dimensions and 0 for the non-compact unwrapped 
directions. This apparent discrepancy arises because we do not perform an 
average of brane wrappings over all dimensions as in our case the wrapped 
dimensions are fixed and topologically distinct from the non-compact 
dimensions. In the conventional BGC one has to perform the average as all 
dimensions are equivalent and then the averaged equation of state parameter 
will indeed be given by
$$\frac{(-1).\Ds+0.D}{\Dh-1}=-\frac{\Ds}{\Dh-1}$$
Before proceeding any further, let us check that the stress energy tensor is 
indeed consistent, i.e. it satisfies the divergence free condition
\be
\n_{\ah}T^{\ah\bh}=0
\label{eq:divergence}
\ee
Now
$$\n_{\ah}T^{\ah\bh}=e_{\ah}T^{\ah\bh}+\om^{\ah}{}_{\ch\ah}T^{\ch\bh}+\om^{\bh}{}_{\ch\ah}T^{\ah\ch}$$
and the non-zero connection co-efficients for (\ref{eq:vielbein}) are given 
by
$$\widehat{\om}^{\au}{}_{0\cu}=\de_{\cu}{}^{\au}e^{-W}\dot{A}$$
$$\widehat{\om}^{\as}{}_{0\cs}=\de_{\cs}{}^{\as}e^{-W}\dot{S}$$
\be
\widehat{\om}^{\as}{}_{\bs\cs}=e^{-S}\om^{\as}{}_{\bs\cs}
\ee
Then trivially
$$\n_{\ah}T^{\ah\bu}=0$$
$$\n_{\ah}T^{\ah 0}= e_{\ah}T^{\ah 0}+\om^{\ah}{}_{\ch\ah}T^{\ch 0}
+\om^{0}{}_{\ch\ah}T^{\ah\ch}$$
$$=e^{-W}\p_t(nT_pe^{-DA})+e^{-W}(D\dot{A}+\Ds\dot{S})nT_pe^{-DA}-e^{-W}nT_p\Ds\dot{S}e^{-DA}=0$$
Finally
$$\n_{\ah}T^{\ah\bs}=e_{\ah}T^{\ah\bs}+\om^{\ah}{}_{\ch\ah}T^{\ch\bs}+\om^{\bs}{}_{\ch\ah}T^{\ah\ch}$$
$$=0+\om^{\as}{}_{\ds\as}T^{\ds\bs}+\om^{\bs}{}_{\ds\as}T^{\as\ds}=0$$
since $\om^{\as}{}_{\ds\as}=0$ for unimodular groups, and compact groups are 
unimodular.
\begin{center} {\bf {\large III. BRANE GAS COSMOLOGY }} \end{center}
\noindent
{\bf Without Matter:} Having obtained the energy momentum tensor for the 
brane gas, we can now proceed
to obtain Einstein's  equations of motion. In order to do that we first 
compute
the Einstein tensor for the geometry 
(\ref{eq:vielbein},\ref{eq:flatmetric}).
$$
G_{00}=e^{-2W}\left[\2 D(D-1)\dot{A}^2+\2
\Ds(\Ds-1)\dot{S}^2+D\Ds\dot{A}\dot{S}+\2\Ds \sto{\la}e^{2(W-S)}\right]$$
$$
G_{\au\bu}=-\de_{\au\bu}e^{-2W}\left[(D-1)\ddot{A}+
\Ds\ddot{S}+\2 D(D-1)\dot{A}^2+\2
\Ds(\Ds+1)\dot{S}^2\right.$$
$$\left.+\Ds(D-1)\dot{A}\dot{S}-(D-1)\dot{A}\dot{W}-\Ds\dot{S}\dot{W}+\2\Ds 
\sto{\la}e^{2(W-S)}\right]$$
$$
G_{\as\bs}=-g_{\as\bs}e^{-2W}\left[D\ddot{A}+
(\Ds-1)\ddot{S}+\2 D(D+1)\dot{A}^2+\2
\Ds(\Ds-1)\dot{S}^2\right.$$
\be
\left.+D(\Ds-1)\dot{A}\dot{S}-D\dot{A}\dot{W}-(\Ds-1)\dot{S}\dot{W}+\2(\Ds-2)\sto{\la}e^{2(W-S)}\right]
\ee
Einstein's field equations then read
\be
G_{\ah\bh}=\kappa^2T_{\ah\bh}
\label{eq:fieldeqns}
\ee
To analyse these equations in more detail it is convenient to choose the 
`canonical gauge'\footnote{This is the gauge where in the action the kinetic 
terms for the variables $S$ and $A$ are canonical, and hence we expect  
Newton's law type equations which provide a clear picture of the dynamics. 
One can always go back to the more standard $W=0$ gauge.} \cite{kaya}:
\be
W=DA+\Ds S
\ee
After some recombinations one obtains
$$
(\Dh-2)\ddot{A}=(\Ds+1)\kappa^2nT_pe^{2\Ds S+DA}
$$
\be
(\Dh-2)\ddot{S}=-(D-2)\kappa^2nT_pe^{2\Ds 
S+DA}-\sto{\la}(\Dh-2)e^{2(\Ds-1)S+2DA}
\ee
The third equation is not expected to be linearly independent as usually 
happens in general relativity, though it can constraint the initial 
conditions.
These equations look the same as with branes wrapping flat tori \cite{kaya}, 
the only addition being  the internal curvature term. However, the 
curvature term has the same sign as that of the brane term and thus it can 
only enhance the brane gas  effect of trying to contract the extra 
dimensions. Unfortunately we couldn't find an analytical solution as it was 
possible in the flat case, but numerical evolution of the equations 
exemplify the behaviour.
%\begin{center}
%\scalebox{.7}{\epsfig{bevolution1.eps,angle=270}}\\
\begin{figure}[t]
\begin{center}
\includegraphics[scale=0.5,angle=270,clip=]{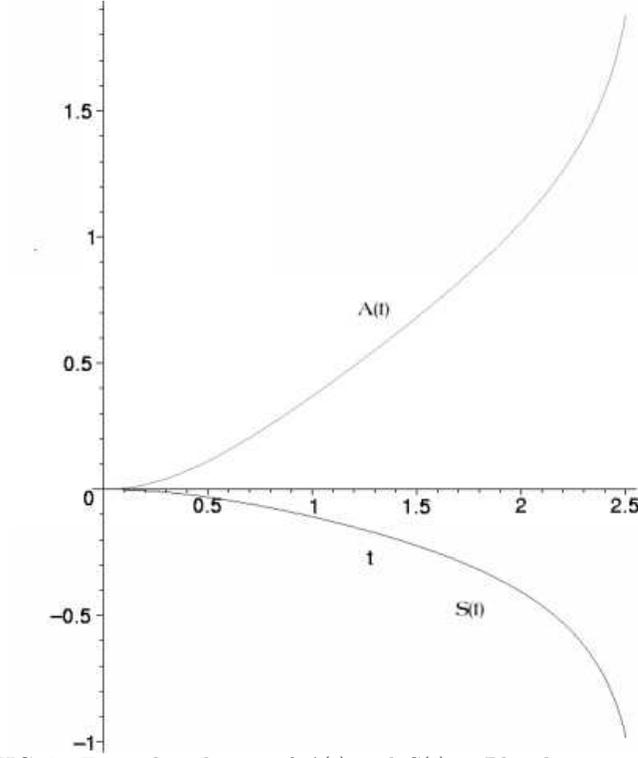}
\caption{Typical evolution of $A(t)$ and $S(t)$ in Planckian units.}
\end{center}
\end{figure}
As expected we find that while the unwrapped dimensions grow, the  wrapped 
dimensions shrink (see figure) thus providing an explanation for the differential size of 
the external and the internal dimensions.
\vs\\
{\bf With Matter:} Let us now make the picture more realistic by adding 
matter/radiation:
$$T^M_{00}=\rho$$
$$T^M_{\au\bu}=\de_{\au\bu}p$$
\be
T^M_{\as\bs}=g_{\as\bs}\sto{p}
\label{eq:matterstress}
\ee
One can solve for the energy density $\rho$ using the equations of state
$$p=\om \rho$$
\be
\sto{p}=\sto{\om} \rho
\label{eq:state}
\ee
and the divergence equation
$$\n_{\ah}T^{\ah 0}=0$$
One obtains exactly the same result as for the flat case:
\be
\rho=\rho_0 e^{-D(1+\om)A-\Ds(1+\sto{\om})S}
\ee
It is easy to check that other equations in (\ref{eq:divergence}) are 
automatically satisfied. It is interesting to note that the brane stress 
energy tensor is also of the same form as 
(\ref{eq:matterstress},\ref{eq:state}) with $\om=0$, $\sto{\om}=-1$ and 
$\rho_0=nT_p$.

The addition of radiation/matter modifies the equations of motion to
$$
(\Dh-2)\ddot{A}=(\Ds+1)\kappa^2nT_pe^{2\Ds 
S+DA}+\kappa^2\rho_0(1+(\Ds-1)\om-\Ds\sto{\om})e^{D(1-\om)A+\Ds(1-\sto{\om})S}
$$
$$
(\Dh-2)\ddot{S}=-(D-2)\kappa^2nT_pe^{2\Ds 
S+DA}-\sto{\la}(\Dh-2)e^{2(\Ds-1)S+2DA}$$
\be
+\kappa^2\rho_0(1+(D-1)\sto{\om}-D\om)e^{D(1-\om)A+\Ds(1-\sto{\om})S}
\label{eq:mattereqns}
\ee
It is clear that the evolution would be complex and there could be several 
regimes depending upon the specific values of the equation of state 
parameters and the values of $S$ and $A$. In general one cannot solve these 
equations analytically and to comprehensively understand the dynamics one 
will have to resort to numerical simulations. However, let us consider a 
likely scenario:

It is desirable to account for an inflationary phase in the beginning. Thus 
one can assume that initially $\rho_0$ comes from a cosmological term (or 
perhaps an effective cosmological term arising from slow rolling of a scalar 
field, with $\rho_0=\rho_{\mt{inflaton}}$) with $\om=\sto{\om}=-1$. It is 
easy to see from the evolution equations that when both $A$ and $S$ are 
small enough then the dominant exponents are associated with the internal 
curvature and the brane contributions and hence the evolution would proceed 
similar to  the case when no matter was present. $A$ will expand while $S$ 
will shrink giving us the origin of the differential size of $A$ and $S$. As 
$A$ increases, there will come a point when cosmological term starts to 
dominate over the brane term. Also note, the internal curvature term 
always dominate over the brane term because $S$ is small while $A$ is large. 
Thus the phase of differential `growth' will soon give way to an 
inflationary phase where the size of the internal manifold remains approximately 
constant while that of the external manifold increases exponentially (this 
is nothing but the familiar de Sitter vacuum). After the end of inflation, 
presumably the universe becomes radiation and later matter dominated:
$$\rho_{\mt{inflaton}}\ra\rho_{\mt{radiation}}\ra\rho_{\mt{matter}}$$

At this point one may wonder as to what kind of equation of state parameter 
does matter and radiation obey along the extra dimensions? Although the 
answer is not clear it seems natural that since the internal manifold is now 
very small  the wave-function of the matter/radiation particles will  wrap 
around the extra-dimensional manifold and thus effectively behave in much 
the same way as branes do, or in other words have $\sto{\om}=-1$. Since by 
this time we expect most of the branes to have annihilated the evolution 
would be governed by a competition between the internal curvature and the 
radiation/matter terms, unlike between brane and radiation/matter as was 
discussed in \cite{kaya} for the flat case.

One can now try to find exact solutions (assuming $T_p=0$) of 
(\ref{eq:mattereqns}) by matching the exponents \cite{anupam,kaya}. We find 
it convenient to directly work in the $W=0$ gauge for this purpose, where 
the independent equations look like
$$
\2 D(D-1)\dot{A}^2+\2
\Ds(\Ds-1)\dot{S}^2+D\Ds\dot{A}\dot{S}+\2\Ds 
\sto{\la}e^{-2S}=\kappa^2\rho_0 e^{-D(1+\om)A-\Ds(1+\sto{\om})S}$$
and
$$
D\ddot{A}+
(\Ds-1)\ddot{S}+\2 D(D+1)\dot{A}^2+\2
\Ds(\Ds-1)\dot{S}^2$$
\be
+D(\Ds-1)\dot{A}\dot{S}+\2(\Ds-2)\sto{\la}e^{-2S}=-\kappa^2\sto{\om}\rho_0 
e^{-D(1+\om)A-\Ds(1+\sto{\om})S}
\ee
To match the exponents we make the following ansatz
\be
A(t)=ln(a_0)+a_1ln(t)\ ;\ S(t)=ln(s_0)+s_1ln(t)
\label{eq:ansatz}
\ee
With (\ref{eq:ansatz}) since $\dot{A},\dot{S},\ddot{A},\ddot{S}$, all go as 
$\sim t^{-2}$ we impose
$$-D(1+\om)a_1-\Ds(1+\sto{\om})s_1=-2$$
and
$$-2s_1=-2$$
\be
\Ra s_1=1 \mx{ and } a_1=\frac{2}{3(1+\om)}
\label{eq:solution}
\ee
for $D=3$ and $\sto{\om}=-1$. For both radiation ($\om=1/3$) and matter 
($\om=0$) this gives the usual scaling laws for the size of the external 
universe, $e^A\sim t^{1/2}$ and $e^A\sim t^{2/3}$ respectively. Matching the 
exponents however still leaves us to satisfy two equations with two unknowns 
($a_0,s_0$). In the flat case with branes and matter/radiation, one could 
indeed find such a solution set, but unfortunately it fails in our case 
(although solutions exist for $\om\ra 0^-$). However, as remarked in 
\cite{kaya} (\ref{eq:solution}) can still be taken to indicate a general 
behaviour of the solutions and we here take a similar  view point. We also 
observe that unlike in the flat case where an interplay between the brane 
contraction and expansion due to pressureless dust could stabilize the 
internal manifold, the internal manifold here grows linearly with time! This is because the interplay of curvature terms and matter gives us a ``potential hill'' rather than a valley. To exemplify this consider the evolution of $S$   in  (\ref{eq:mattereqns}) with $T_p=0$ and in the presence of  a pure cosmological constant\footnote{Note that in our approach the equation of state for matter along the extra dimensions is the same as that of a pure cosmological constant term.}. In the equation of motion of  for $S$  the $A$ dependence  factorizes and  the evolution  is governed by an effective potential of the form
\be
V_{\mt{eff}}(S)\sim C_{\mt{curv}}e^{2(\Ds-1)S}
-C_{\mt{cos/mat}}e^{2\Ds S}
\ee
Clearly there is an unstable potential hill and because of the cosmological/matter term $S$ tends to grow once to the right hand side of the hill. This 
suggests that one perhaps needs to incorporate other forms of stabilization 
mechanism, for example by turning on the flux fields \cite{flux}, to be able 
to stabilize the internal manifold when it is curved.
\begin{center}  {\large {\bf SUMMARY AND FUTURE RESEARCH}} \end{center}

In this paper we have tried to describe a framework to study BGC with branes 
wrapping curved group manifolds rather than flat tori which has been 
extensively studied. We find that although at earlier times the dynamics is 
similar to the flat case, the branes now reinforced by the 
internal curvature effects again tries to contract the wrapped compact 
dimensions, the later dynamics is different as the internal curvature terms 
become important. In particular, there seems to be a regime when the 
competition between the internal curvature terms  and the usual 
matter-radiation contributions allow the external universe to scale in the 
usual way $\sim t^{1/2}$ for radiation and $\sim t^{2/3}$ for matter at 
least approximately. However the analysis carried out here is only 
qualitative and to have a better understanding of the dynamics one needs to 
perform a thorough numerical calculation.

One can also try to explore whether one can generate inflation\cite{inflation}  in these 
models without any explicit $\rho_{\mt{infl}}$. Our preliminary numerical 
analysis does seem to suggest (see figure) that the external universe 
undergoes an accelerated phase of expansion at early times while the 
internal manifold is contracting. Such an inflationary regime will come to 
an end when branes annihilate to produce reheating. A related issue would be 
to study the stability of the branes in the first place. This may reveal 
that a brane gas cannot be supported for some of the group manifolds and 
thus their smallness today cannot be explained in the BGC framework.

Finally, it would be interesting to try and compute the annihilation rates 
of branes wrapping different subgroups (or coset spaces) of the internal 
group  manifold and then apply the same reasoning as was done for the tori 
to see if one can generate a hierarchy of size (or squashing) within the 
full group  manifold. It would be even more interesting to be able to start 
out with a non-compact simple group manifold and carry out this analysis to 
see whether the non-compact directions really become large while the compact 
ones are constrained by branes. This would then at least be able to explain 
why today we observe a product structure of the external and the internal 
universe. Of course this still will not explain the dimensionality of the 
observed universe.
\vspace{5mm}\\
{\bf Acknowledgements:} I would like to  thank Horace Stoica for helpful discussions and suggestions. This work is supported  by the Natural Sciences and Engineering Research Council of Canada, Grant No. 204540.

\end{document}